\documentstyle[apj,times,psfig]{article}

\def\beq{\begin{equation}} \def\eeq{\end{equation}}
\def\ie{i.e$.$~} \def\eg{e.g$.$~} \def\eq{eq$.$~}  \def\etal{et al$.$~}
\def\epsel{\varepsilon_e} \def\epsmag{\varepsilon_B} \def\cm3{\;{\rm cm^{-3}}} \def\deg{^{\rm o}}
\def\epse1{\varepsilon_{e,-1}} \def\epsB2{\varepsilon_{B,-2}} \def\E053{{\cal E}_{0,53}}
\def\simg{\mathrel{%
      \rlap{\raise 0.511ex \hbox{$>$}}{\lower 0.511ex \hbox{$\sim$}}}}
\def\siml{\mathrel{%
      \rlap{\raise 0.511ex \hbox{$<$}}{\lower 0.511ex \hbox{$\sim$}}}}

\begin{document}

\title{Properties of Relativistic Jets in Gamma-Ray Burst Afterglows}

\author{A. Panaitescu}
\affil{Dept. of Astrophysical Sciences, Princeton University, Princeton, NJ 08544}
\and
\author{P. Kumar}
\affil{Institute for Advanced Study, Olden Lane, Princeton, NJ 08540}

\begin{abstract}

 We extend our calculation of physical parameters of GRB jets by modeling the broadband 
emission of the afterglows 970508, 980519, 991208, 000926, 000418, and 010222. Together 
with 990123, 990510, 991216, and 000301c, there are ten well-observed afterglows for which 
the initial opening angle of the GRB jet can be constrained. The jet energies (after the 
GRB phase) obtained for this set of afterglows are within one decade around $5\times 10^{50}$ 
erg. With the exception of 000418, which requires a jet wider than 1/2 radians, the jet 
initial half-angle in the other cases ranges from $2\deg$ to $20\deg$. 
We find that, in half of the cases, a homogeneous ambient medium accommodates the afterglow 
emission better than the wind-like $r^{-2}$ profile medium expected around massive stars. 
The two types of media give fits of comparable quality in four cases, a wind medium providing 
a better description only for 970508. The circumburst densities we obtain are in the 
$0.1-100\cm3$ range, with the exception of 990123, for which it is below $10^{-2}\cm3$. 
If in all ten cases the observed GRB durations are a good measure of the ejecta deceleration 
timescale, then the parameters obtained here lead to jet initial bulk Lorentz factors 
between 70 and 300, anticorrelated with the jet initial aperture, and jet masses around 
$10^{-6} M_\odot$. Our results on the jet energy, opening Lorentz factor, and evacuation 
of material until break-out provide constraints on theoretical models of GRB jets.

\end{abstract}


\section{Introduction}

 The localization of Gamma-Ray Bursts (GRBs) to within a few arc-minutes by the Italian--Dutch 
satellite BeppoSAX, the Interplanetary Network, and the Rossi--$X$-ray Transient Explorer have 
enabled us to carry out ground-based follow-up searches for afterglow emission. The current 
database of multiwavelength (radio, millimeter, optical, and $X$-ray) observations allows us 
to begin a statistical study of the physical properties of GRB afterglows.

 This is a third in a series of papers modelling the broadband emission of GRB afterglows, with
the aim of determining the total energy in the relativistic ejecta, the jet opening angle, the 
density and profile of the medium in the immediate vicinity ($\siml 10^{18}$ cm) of the burst,
and the microphysical shock parameters. The first paper (PK01) has presented the modelling of 
the afterglows 990123, 990510, and 991216 while the second (Panaitescu 2001) analyzed the 
peculiar afterglow 000301c, whose emission fall-off exhibited a sharp break. Here we present 
our results for the afterglows 970508, 980519, 991208, 000926, 000418, and 010222. With the 
exception of 000418, after 1 day the decay of their optical emission is steep or exhibits a 
steepening, as expected if the GRB ejecta are well collimated (Rhoads 1999), thus their modelling 
allows the determination of the jet aperture and energy. Section \S\ref{model} outlines the 
model used to fit the broadband emission of these afterglows. The jet properties inferred for 
individual afterglows are presented in \S\ref{numeric} and the results for the entire set are 
analyzed in \S\ref{features}. 

\vspace*{1cm}

\section{The Afterglow Model}
\label{model}

\subsection{Model Features}

 The calculation of the afterglow emission is carried out in the standard framework of relativistic
ejecta decelerated by an external medium (M\'esz\'aros \& Rees 1997), with allowance for the
effects due to collimation (Rhoads 1999). The equations governing the dynamics of jet--medium 
interaction and those for the calculation of the synchrotron and inverse Compton emission are 
presented in KP00, PK00 and PK01. Similar analytical 
treatments of jet dynamics and/or emission of radiation can be found in Waxman (1997), Granot, 
Piran \& Sari (1999), Gruzinov \& Waxman (1999), Wijers \& Galama (1999), Chevalier \& Li (2000), 
and Sari \& Esin (2001). The effect of interstellar scintillation on the radio 
afterglow emission (Goodman 1997) is taken into account following the treatment of Walker (1998).

 In our treatment, the afterglow modelling has the following features: \\
$1)$ the jet is considered uniform, with an energy per solid angle independent of direction within
     the jet; \\
$2)$ the shocked gas internal energy density is assumed uniform; \\
$3)$ the jet dynamics is calculated by following the evolution of its energy (which decreases 
     due to radiative losses), mass (increasing, as the jet sweeps-up the surrounding medium), 
     and aperture (which increases due to jet expansion in the comoving frame). The equations
     employed are accurate in any relativistic regime; \\
$4)$ the afterglow emission is calculated by integrating over the jet dynamics the synchrotron and 
     inverse Compton radiation, taking into account the spread in the arrival time of photons emitted
     at a given radius. The Compton parameter and its evolution are calculated from the electron
     distribution; \\
$5)$ the shock-accelerated electron distribution is assumed a power-law ${\cal N}(\gamma) \propto 
     \gamma^{-p}$ in the random electron Lorentz factor $\gamma$, starting from a minimum $\gamma_i$ 
     up to a high energy break $\gamma_*$; \\
$6)$ the fractional energy in electrons and magnetic field are constant throughout the jet 
     deceleration; \\
$7)$ in general, the observer is assumed to lie on the jet axis. For our jet model, observer
     off-sets lower than the jet initial aperture produce insignificant changes in the resulting
     light-curve (see Granot \etal 2002), and therefore in the fitting parameters.

\subsection{Model Parameters and Their Determination}

 The model has {\it three} parameters that give the jet dynamics: the initial jet energy $E_0$, 
initial half-angle $\theta_0$, and external particle density $n$ (or the constant $A$ for a wind-like 
density profile\footnote{
 This constant is proportional to the ratio between the mass loss rate of the star which ejects
 the wind and the speed of this wind. We denote by $A_*$ the value of the constant $A$
 relative to that corresponding to $10^{-5}\, M_\odot$ ejected per year at a speed of 1000 km/s.}
$n(r) = A r^{-2}$), and {\it three} parameters related to the microphysics of shocks: 
the fraction $\epsmag$ of the post-shock energy density in magnetic fields, the fractional energy 
$\epsel$ in electrons if they all had the same Lorentz factor $\gamma_i$, and the power-law index 
$p$. For $p \siml 2$, there are {\it two} additional parameters: the fractional energy $\epsilon$ 
of the electrons between $\gamma_i$ and $\gamma_*$, which parameterizes $\gamma_*$, and the power-law 
index $q > 2$ of the electron distribution above $\gamma_*$ (for simplicity the cut-off above 
$\gamma_*$ is assumed to be a power-law). In some cases, the curvature of the optical spectrum
or consistency between the optical and $X$-ray afterglow emission indicates a significant dust
extinction in the host galaxy. This is taken into account by assuming an SMC-like reddening
curve and adding an extra model parameter, the $A_V$ extinction in the host rest-frame. We note
that the initial jet Lorentz factor has little effect on the afterglow emission and is not
considered a free model parameter.

 The spectrum of the afterglow synchrotron emission (Sari, Piran \& Narayan 1998) has breaks at the 
self-absorption frequency $\nu_a$, injection frequency $\nu_i$ corresponding to the minimum electron 
$\gamma_i$, cooling frequency $\nu_c$ corresponding to the electron Lorentz factor for which the 
radiative timescale equals the dynamical time, and cut-off frequency $\nu_*$ associated with 
$\gamma_*$. Generically, the afterglow emission $F_\nu$ can be written as
\beq
 F_\nu = F_p \;\nu_a^{-\beta_a} \nu_i^{-\beta_i} \nu_c^{-\beta_c} \nu_*^{-\beta_*} \;
         \nu^{\beta_a+\beta_i+\beta_c+\beta_*} \;,
\label{Fnu}
\eeq
where the exponent $\beta$ is non-zero for all break frequencies between the observing frequency
$\nu$ and the frequency $\nu_p = \min (\nu_c, \nu_i)$ at which the synchrotron spectrum peaks, 
$F_p$ being the flux at $\nu_p$. For a relativistic spherical outflow and an adiabatic expansion, 
the break frequencies and peak flux are given by:
\beq
 \left. \begin{array}{ll}
 \nu_a \sim 2\; (z+1)^{-1}\; \E053^{1/5}\; n_0^{3/5}\; \epse1^{-1}\; \epsB2^{1/5} & {\rm GHz} \\
 \nu_i \sim 20\; (z+1)^{1/2}\; \E053^{1/2}\; \epse1^2\; \epsB2^{1/2}\;  t_d^{-3/2} &  {\rm THz} \\
 \nu_c \sim 600\; (z+1)^{-1/2}\; \E053^{-1/2}\; n_0^{-1}\; \epsB2^{-3/2}\; t_d^{-1/2} & {\rm THz} \\
 F_p  \sim 20\; (z+1) D_{L,28}^{-2}\; \E053\; n_0^{1/2}\; \epsB2^{1/2}  & {\rm mJy}
  \end{array} \right. \;
\label{spectrum}
\eeq
where $z$ is the afterglow redshift, $\E053$ is the fireball energy in $10^{53}$ erg, $n_0$ the 
external medium density y in $\cm3$, $\epsel$ and $\epsmag$ have been normalized to 0.1 and 0.01 
respectively, $t_d$ is the observer time measured in days, and $D_{L,28}$ is the burst luminosity 
distance measured in $10^{28}$ cm. Similar equations can be derived for spreading jets, 
non-relativistic GRB remnants, or non-adiabatic expansion.

 As illustrated in equation (\ref{Fnu}), the afterglow light-curve at a given frequency is given 
by the evolution of $F_p$, $\nu_a$, $\nu_i$, $\nu_c$, and $\nu_*$ which, for constant parameters 
$\epsmag$, $\epsel$ and $\epsilon$, are determined by the Lorentz factor $\Gamma$ of the jet, 
its radius $r$, and external density profile $n(r)$. For a relativistic jet and negligible radiative 
losses, conservation of total jet energy leads to $\Gamma \propto t^{-3/8}$ and $r \propto t^{1/4}$ 
in the case of a homogeneous medium and $\Gamma \propto t^{-1/4}$, $r \propto t^{1/2}$ for a wind 
external medium, {\it before} the time
\beq
 t_j = 0.4\;(z+1) \left( E_{0,50} n_0^{-1} \right)^{1/3} \theta_{0,-1}^2 \;{\rm day}\;,
\label{tjet}
\eeq
when the jet transits between a quasi-spherical expansion and a lateral spreading dominated 
one \footnote{
     This is also the time when the jet Lorentz factor equals the reciprocal of the jet half-angle,
     thus the radiation emitted from the jet edge is no longer relativistic beamed away from the
     observer.}. 
In equation (\ref{tjet}) $E_{0,50}$ the initial jet energy measured in $10^{50}$ erg and $\theta_{0,-1}$ 
the initial jet half-opening measured in 0.1 radians. {\it After} $t_j$ the jet dynamics is described 
by $\Gamma \propto t^{-1/2}$ and (to ``zeroth order") $r \sim$ constant. The resulting evolution of 
the afterglow spectral characteristics in these two asymptotic regimes are summarized in Table 1, 
together with the afterglow light-curve $t^{-\alpha}$ at frequencies above $\nu_i$, for slowly 
cooling electrons, \ie $\nu_i < \nu_c$, and for Compton parameter $Y$ below or above 1, \ie electron 
cooling due to synchrotron losses or up-scatterings, respectively. 

 Because the afterglow light-curve decay depends only on the index $p$ of the electron distribution 
(or $q$ above $\gamma_*$), this index can be easily determined from observations {\it if} one knows
the location of $\nu_i$ and $\nu_c$ relative to the observing frequency. For measurements made more 
than a few hours after the GRB, the injection frequency is below the optical domain, thus the only 
uncertainties are related to $\nu_c$. Consistency between the decay indices $\alpha (p)$ given in 
Table 1 and the slope $\beta(p)$ of the synchrotron power-law optical spectrum $F_\nu \propto 
\nu^{-\beta}$, where
\beq
 \beta = \frac{1}{2}\; (p-1) \; {\rm for} \; \nu_i< \nu < \nu_c \;, \quad
 \beta = \frac{p}{2}\;  \; {\rm for} \; \nu_c < \nu \;,
\label{beta}
\eeq
is commonly used to determine both $p$ and the location of $\nu_c$ relative to the optical domain.

 Finding the remaining parameters, five if the high frequency cut-off $\nu_*$ is above the highest 
observing frequency, seven in the opposite case, is conditioned by the localization of the spectral 
breaks at some time (not necessarily the same for all breaks), either from the afterglow flux at 
two frequencies bracketing a given break, or from the passage of that break through an observing 
band. If we know $\nu_a$, $\nu_i$, $\nu_c$ and $F_p$ from observations, inverting the set of 
equations (\ref{spectrum}) above allows the calculation of ${\cal E}_0$, $n$, $\epsel$ and 
$\epsmag$. Then, if the afterglow decay exhibited a steepening which can be identified as the
``jet-break", equation (\ref{tjet}) gives the initial jet aperture $\theta_0$. Together with 
${\cal E}_0$, this allows the calculation of the jet initial energy $E_0$. 

 In general the above method cannot be readily used to other afterglows because the locations of 
$\nu_a$ and $\nu_c$ are not sufficiently constrained by the available data. For instance, evidence 
for self-absorption at radio frequencies exists only for the afterglows 970508, 991208, 000301c and, 
perhaps, 991216. Furthermore, the approximations usually made in analytical treatments of the 
afterglow emission (\eg Waxman 1997, Wijers \& Galama 1999, PK00, Sari \& Esin 2001) 
are accurate only over a limited time interval. Various departures from those 
approximations, such as $1)$ moderately relativistic jets, with $\Gamma$ of several, $2)$ jets 
transiting between quasi-collimated and lateral-spreading expansion, $3)$ electron radiative 
cooling not dominated by a single emission process (synchrotron or inverse Compton), $4)$ afterglow 
spectral breaks smoothed by the differential relativistic boost and photon arrival-time over the 
jet surface, $5)$ time changing ordering of the spectral, during the afterglow evolution, require
numerical calculations to yield a more reliable determination of jet parameters.

\section{Collimated Afterglows}
\label{numeric}

 The model outlined above was used to model the broadband emission of ten afterglows to determine 
the parameters $E_0$, $\theta_0$, $n$ (or $A_*$ for a wind), $\epsel$, $\epsmag$, and $p$ (plus 
$A_V$, $\epsilon$ and $q$ in those cases where they are relevant) by $\chi^2$-minimization. 
Nine of these afterglows (970508, 980519, 990123, 990510, 991208, 991216, 000301c, 000926, and 
010222) were selected based on the existence of a break in (or a steep decay of) the optical 
light-curve, allowing the calculation of the jet initial opening, and sufficient broadband 
observations to make the modelling meaningful. The 000418 afterglow has been added due to its
good multiwavelength coverage, although its emission does not exhibit a signature of a jet-break. 

 In calculating the afterglow optical fluxes we assumed a 5\% error in the magnitude-to-flux 
conversion and Galactic reddening and we subtracted the reported contributions of the host or 
contaminating galaxies. $X$-ray fluxes have been calculated from the reported band fluxes 
(2--10 keV, usually) and $X$-ray spectral slopes.
 
 The modelling of the afterglows 990123, 990510, 991216, and 000301c is presented elsewhere (PK01, 
Panaitescu 2001). The use of a larger set of $X$-ray measurements for the 990123 afterglow (Costa 
1999) and the reduction of the assumed error in the Galactic extinction of the optical emission of 
991216 from 10\% to 5\%, led in these two cases to small changes in the best fit parameters. Below
we discuss the features of the broadband emission and the modelling of the remaining six afterglows.

\subsection{GRB 970508}

 During the first day after the burst, the optical emission of 970508 exhibits a brightening 
by more than one magnitude, lasting for about 1 day, followed by a long-lived decay of temporal
index $\alpha_o = 1.17 \pm 0.03$ (Sokolov \etal 1998). Simultaneous with the brightening,
Galama \etal (1998b) observe a reddening of the optical spectrum, the slope changing from
$\beta_o \sim 0.5$ at 1 day to $\beta_o \sim 1.1$ at $t > 2$ days. They interpret the softening
of the optical spectrum as the passage of the cooling frequency $\nu_c$, thus the latter 
$\beta_o$ implies $p = 2.2$ .

 The radio emission of this afterglow has been monitored for about 1 year. Frail, Waxman \& 
Kulkarni (2000b) report that the slope of the radio spectrum evolved from $\beta_r = -0.25
\pm 0.04$ before 80 days to an average $\beta_r = 0.6 \pm 0.2$ after 100 days, which must be
due to the passage of the injection frequency $\nu_i$. Therefore the latter $\beta_r$ implies
that $p = 2.2 \pm 0.4$, as the radio domain is expected to lie below $\nu_c$. After 100 days,
the radio emission decayed as a power-law of index $\alpha_r = 1.2 \pm 0.1$ (Frail \etal 2000b),
which, given the above $p$ and that $\alpha \sim p$ after the jet-break time, suggests that 
the $t_j \simg 100$ days, if the assumption of relativistic motion is still correct.
 
 In order to have an observational constrain on the jet initial opening, we make the assumption
that the brightening seen at 1 day is due to an observer location $\theta_{obs}$ which, initially, 
is outside the jet (\ie $\theta_{obs} > \theta_0$). In this case, the observer will see a rising 
light-curve when the jet has decelerated down to a Lorentz factor $\Gamma \sim (\theta_{obs} -
\theta)^{-1}$, for which the radiation from the nearest part of the jet becomes visible to the
observer. We also assume that the GRB and the dimmer afterglow emission prior to the brightening 
were due to some less energetic ejecta located outside the main jet and moving toward the
observer, but ignore their effect on the dynamics of the jet and afterglow emission after
1 day.

 The parameters of the best fit to the radio (Frail \etal 2000b), millimeter (Bremer \etal 1998),
optical (Sahu \etal 1997, Galama \etal 1998a, Pedersen \etal 1998, Sokolov \etal 1998, Zharikov 
\etal 1998), and $X$-ray (Piro \etal 1998) emission of 970508 after the onset of the brightening 
phase, obtained with a {\sl homogeneous} external medium, are given in Table 2. The observer is 
located at $\theta_{obs} \sim 4/3\, \theta_0$. The fit is not satisfactory (thus we do not
determine confidence intervals), exhibiting a slower brightening and weaker $X$-ray emission than 
observed. 
The jet isotropic equivalent energy ${\cal E}_0$ derived by Wijers \& Galama (1999) for 
the 970508 afterglow from its spectrum at 12 days is twice smaller than our value\footnote{ 
   For the $\epsel$ and $\epsmag$ given for 970508 in Table 2, radiative losses 
   are significant (80\% until around 10 days), thus the jet energy we infer is likely to 
   exceed that obtained by other researchers using adiabatic models for the jet dynamics}
while their value for $\epsmag$ is twice larger. The largest discrepancies are in $n$, our value 
being 25 times larger than theirs, and in $\epsel$ for which we find a value five times larger.
From the same 12 day spectrum of 970508, Granot \etal (1999) found ${\cal E}_0$ 15 times smaller, 
$n$ four times larger, and $\epsmag$ three times smaller than our values, and a similar $\epsel$.
Through an analytical treatment of the radio emission of 970508, Frail \etal (2000b) inferred a 
jet energy 4 times smaller than obtained by us, electron and magnetic parameters close to 
equipartition and $n \sim 1\cm3$, close to our results.  

 A significantly better fit can be obtained for a wind-like external medium. This fit, shown
in Figure 1, has $E_0 = 1.6 \times 10^{51}$ erg, $\theta_0 = 18\deg$, $\theta_{obs} \sim 5/3\,
\theta_0$, $A_* = 0.39$, $\epsel = 0.15$, $\epsmag=0.10$, $p = 2.32$, and $\chi^2 = 570$ for 279 
degrees of freedom (dof). Although it describes well the rise of the optical emission, it does 
not accommodate the $X$-ray emission during the brightening. Using a spherical model, Chevalier 
\& Li (2000) estimated similar values for $A_*$ and $\epsmag$, and ${\cal E}_0$ and $n$ twenty
and seven times smaller, respectively.

 The spectral properties of the best fit models discussed above are similar. In both cases the
injection frequency $\nu_i$ is slightly below the optical domain at 1 day\footnote{
   Due to the off-axis location of the observer, the passage of $\nu_i$ through
   the optical occurs later than predicted by \eq (\ref{spectrum})}
, its evolution yielding the spectral softening observed by Galama \etal (1998b). 
At $\siml 100$ days, $\nu_i$ passes through the radio domain, as inferred from observations by 
Frail \etal (2000b).

\subsection{GRB 980519}

 The optical emission of this afterglow had a break of magnitude $\Delta \alpha \simeq 0.5$ at 
$t \sim 1$ day, with a temporal index $\alpha_o = 2.22 \pm 0.04$ (Jaunsen \etal 2001) after the
break, close to that measured in $X$-rays, $\alpha_x = 2.25 \pm 0.04$, at about 1 day (Nicastro
\etal 1999). The equality of these two indices is consistent with the achromatic break expected
in the jet model. At $t \siml 1$ day the slope of the optical spectrum dereddened for Galactic 
extinction, $\beta_o = 1.20 \pm 0.25$ (Halpern \etal 1999), is shallower than that measured by 
Nicastro \etal (1999) at about the same time in $X$-rays, $\beta_x = 1.72 \pm 0.42$. The difference 
between the two slopes is close to that expected when $\nu_c$ is between optical and $X$-rays but, 
given the their large uncertainties, does not provide a compelling proof. 

 Numerically we find that the radio (Frail \etal 2000a), optical (Vrba \etal 1999, Jaunsen \etal 
2001) and $X$-ray (Nicastro \etal 1999) emission of 980519 can be well accommodated by a spreading 
jet interacting with a homogeneous medium, and with $\nu_c$ between optical and $X$-rays (Figure 2). 
The best fit with a jet interacting with wind medium has $E_0 = 1.1 \times 10^{51}$ erg, 
$\theta_0 = 6.7\deg$, $A_* = 3.5$, $\epsel = 0.036$, $\epsmag=0.22$, $p = 2.43$, and $\chi^2 = 73$ 
for 46 dof. This model yields a shallower break than observed in the $I$-band light-curve of 
980519 and provides a poorer fit to the radio data.

\subsection{GRB 991208}

 The radio emission of this afterglow (Galama \etal 2000) exhibited a quasi-flat behavior 
until $\sim 10$ days (Figure 3), followed by a power-law decay (Galama \etal 2002) which is 
much shallower than the $t^{-2.2\pm0.2}$ observed in the optical at 2--7 days (Castro-Tirado 
\etal 2001).  The flatness of the early radio emission indicates that the $\nu_i$ frequency is 
above the radio domain and that either the external medium is homogeneous and the GRB remnant 
is a jet observed after $t_j$ (\ie after the jet-break) or the ambient medium is wind-like and 
the GRB jet is seen before $t_j$. This suggests that the steepening of the radio emission at 
10 days is due to the $\nu_i$-passage. Then the shallow radio decay after 10 days requires a 
hard electron distribution, $p < 2$, and the steep optical decay implies that the $\nu_*$-break 
is below the optical domain.  

 The best fit to the data obtained with a jet interacting with a homogeneous medium is shown in 
Figure 3 and has the parameters given in Table 2. The spectral characteristics (break frequencies 
and peak flux) of the model afterglow emission are consistent with those obtained by Galama \etal 
(2000) by fitting the spectrum of 991208 at four epochs. The electron distribution cut-off at
higher energy is characterized $\epsilon = 0.49$ and $q = 2.7$.

 The data can be fit equally well ($\chi^2 = 110$ for 97 dof) with wind medium of $A_* = 0.65$ 
and a jet with parameters $E_0 = 3.2 \times 10^{50}$ erg, $\theta_0 = 14\deg$, $\epsel = 0.054$, 
$\epsmag=0.021$, $p \sim 1.4$, $\epsilon = 0.32$, and $q = 3.1$. The last five parameters above
and the implied ${\cal E}_0$ are similar to those determined by Li \& Chevalier (2001).

\subsection{GRB 000418}

 The $R$-band emission of this afterglow exhibited a flattening after only 10 days (Klose \etal
2000, Berger \etal 2001), indicating an underlying host galaxy. By fitting the optical light-curve
with the power-law fall-off expected of GRB afterglows plus the constant contribution of
the host, Berger \etal (2001) obtain a decay index $\alpha_o = 1.41 \pm 0.08$. The 8.5 GHz radio 
emission exhibits a gradual steepening to a power-law of similar index, $\alpha_r = 1.37 \pm 0.10$
(Berger \etal 2001) after $\sim 40$ days, which, most naturally, is due to the $\nu_i$ frequency
falling below the radio domain. The emission of the 000418 afterglow does not exhibit a steepening
that can be attributed to the spreading of a jet, therefore its modeling should not constrain
significantly the initial jet aperture. 

 Optical observations at other frequencies are rather scarce, making the determination of the
optical spectral slope rather difficult. From the only two simultaneous $K$ and $R$ measurements
at times when the afterglow emission is dominant, we infer $\beta_o = 1.62 \pm 0.15$ (note that
Klose \etal 2000 find a redder spectrum, with $\beta_o = 1.90 \pm 0.15$). Then consistency between
$\alpha_o$ (Table 1) and $\beta_o$ (\eq [\ref{beta}]) suggests that the cooling frequency
is below the optical range and/or there is a significant dust reddening within the host galaxy.

 With our jet model, the best fit for this afterglow is obtained with a wide jet ($\theta_0 
\sim$ 1 radian) interacting with either a homogeneous medium or a wind. Table 2 lists the 
parameters in the former case, the model light-curves being shown in Figure 4. In the latter 
case, the best fit parameters are $E_0 = 2.2 \times 10^{51}$ erg, $\theta_0 = 60\deg$, $A_* = 
0.69$, $\epsel = 0.10$, $\epsmag=0.027$, and $p = 2.04$. The two models have $\chi^2 = 55$ and
$\chi^2 = 56$, respectively, for 61 dof. Taking the host extinction $A_V$ as a free parameter,
does not improve the fits significantly. Compared to the parameters inferred by Berger \etal 
(2001), the above $n$ and $A_*$ are larger by a factor 100 and 10, respectively, $\epsel$ is 
twice larger, $\epsmag$ is smaller by a factor 10-30, and the isotropic ${\cal E}_0$ for the 
$E_0$ and $\theta_0$ above is smaller by a factor of a few. We note that jets narrower than 
1/3 radians yield poorer fits, with $\chi^2 \simg 130$ for 61 dof, while acceptable fits can 
be obtained with spherical ejecta.

\subsection{GRB 000926}

 The $X$-ray emission (Piro \etal 2001) of this afterglow provided for the first time evidence 
(Harrison \etal 2001) that the $X$-ray emission may be inverse Compton scatterings (PK00, Sari 
\& Esin 2001). This is suggested by that the extrapolation of the optical spectrum, after 
dereddening for the host (intrinsic) extinction, falls below the observed $X$-ray flux. 

 The optical emission of 000926 exhibited a break of magnitude $\Delta \alpha \sim 0.75$ at
few days, with a post-break temporal index $\alpha_o \sim 2.35 \pm 0.05$ (Fynbo \etal 2001, 
Price \etal 2001). If interpreted as a jet break, it requires that $p \siml 2.4$, which
would imply (\eq[\ref{beta}]) an optical spectrum significantly harder than observed at $t \sim 1$ 
day: $\beta_o = 1.42 \pm 0.06$ (Fynbo \etal 2001) or $\beta_o = 1.53 \pm 0.07$ (Price \etal 2001).
Within the fireball model, consistency between the optical spectral slope and temporal index 
requires a significant host extinction. From the curvature of the near infrared--optical spectrum, 
Fynbo \etal (2001) infer $A_V = 0.18 \pm 0.06$, corresponding to an extinction in the observer
$I$-band of $0.4 \pm 0.1$ magnitudes, thus the dereddened afterglow spectrum has an optical slope 
$\beta_o \sim 1$. Then equation (\ref{beta}) and $p \sim 2.3$ imply that $\nu_c$ is below the 
optical domain. 
 
 For a homogeneous medium, the best fit obtained with a model with the above features (Figure 5) 
has parameters (table 2) that are close to those obtained Harrison \etal (2001), except $E_0$ 
and $\epsmag$, for which we find values three times smaller and eight times larger, respectively. 
The best fit model with a wind medium has $\chi^2 = 270$ for 102 dof, yielding radio 
fluxes larger than observed, and parameters $E_0 = 2.7 \times 10^{51}$ erg, $\theta_0 = 2.0\deg$, 
$A_* = 2.0$, $\epsel = 0.042$, $\epsmag = 1.6 \times 10^{-4}$, and $p = 2.70$. Note, however, that 
Harrison \etal (2001) found a significantly better fit ($\chi^2 = 167$ for 114 data points) for a 
wind medium.

\subsection{GRB 010222}

 The index $\alpha$ of the power-law decay of the optical emission of 010222 steepened by 
$\Delta \alpha = 0.6 \pm 0.1$ at about 0.5 days, to $\alpha_o = 1.30 \pm 0.05$ (Masetti \etal 2001,
Stanek \etal 2001), a value similar to that seen in the $X$-rays, $\alpha_x = 1.33 \pm 0.04$ 
('t Zand \etal 2001). 
The jet interpretation of this break requires an electron distribution with $p \sim 4/3$. 
As in the case of 991208, a hard electron distribution lowers the cut-off frequency $\nu_*$
sufficiently to yield a break of the afterglow decay when $\nu_*$ passes through of the
observing band. This provides a natural explanation for the second steepening observed in the 
optical after 10 days by Fruchter \etal (2001) (see also http://www.stsci.edu/~fruchter/GRB/010222).

 The low index $p$ required by the jet interpretation of the first break also implies an intrinsic 
optical spectrum harder than $p/2 \sim 2/3$ , \ie harder than observed: $\beta_o = 0.89 \pm 0.03$ 
(Jha \etal 2001, Lee \etal 2001), indicating the existence of a significant dust reddening in the 
host galaxy. The best fit we found to the radio (Berger \& Frail 2001), optical (Cowsik \etal 2001,
Masetti \etal 2001, Sagar \etal 2001, Stanek \etal 2001) and $X$-ray ('t Zand \etal 2001) data has 
$A_V = 0.21$ and a large $\chi^2 = 236$ for 87 dof. More than half of this $\chi^2$ arises from 8 
optical and $X$-ray data, suggesting either that some reported observations have underestimated
uncertainties or that there are short timescale fluctuations in the afterglow emission (Cowsik
\etal 2001). The jet model presented in Figure 6 faces also another problem: even if electrons 
acquire 100\% of the post-shock fluid energy, the passage of $\nu_*$ through the $X$-ray domain 
takes place too early, at about 1 day, leading to a discrepancy between the last few $X$-ray data 
and the model expectations. Given the unsatisfactory fit provided by this model, we do not include 
in Table 2 the uncertainties of the best fit parameters.  

 As shown in Figure 6, the millimeter model emission falls below the detections reported by Fich 
\etal (2001) and Kulkarni \etal (2001) at 220 GHz and 350 GHz, and below the upper limit on the 
95 GHz emission found by Bremer \etal (2001). The constancy of the observed fluxes over almost one 
decade in time and the steep spectrum between 220 GHz and 350 GHz are hard to accommodate within 
the jet model, suggesting that the millimeter excess seen in 010222 is due to a dusty, star-forming 
host galaxy (Kulkarni \etal 2001). These data were not included in our fits.
 
 The best fit obtained with a wind-like medium has a similar large $\chi^2 = 241$ for 87 dof, and 
parameters $E_0 = 2.5 \times 10^{50}$ erg, $\theta_0 = 3.4\deg$, $A_* = 0.18$, $\epsel = 1.4 \times 
10^{-2}$, $\epsmag = 1.0 \times 10^{-3}$, $p = 1.43$, and $A_V=0.22$. It provides a better description 
of the last $X$-ray measurements than the homogeneous medium model, but a poorer one for the decay
steepening seen in the optical at 0.5 days, as jets interacting with winds yield long-lived, smooth
light-curve breaks (KP00). The millimeter emission in the wind model is also consistent with the 
95 GHz upper limit and falls below the 220 GHz and 350 GHz detections.

\section{Jet Properties}
\label{features}

\subsection{Jet Energy}

 For eight of the ten GRBs jets whose basic parameters are listed in Table 2, the ejecta kinetic 
energy $E_0$ at the beginning of the afterglow phase is between $10^{50}$ and $5 \times 10^{50}$ erg. 
A narrow distribution of the jet kinetic energy has also been inferred by Piran \etal (2001) for 
a larger set of afterglows, based on the width of the observed $X$-ray luminosity distribution at 
0.5 days. Note, however, than the jet energy for the 970508 and 000418 afterglows is significantly
larger, being around $2\times 10^{51}$ ergs. 

 The initial half-angle of the jet $\theta_0$ is 
correlated with $E_0$: excluding 000418, for which we find $E_0$ and $\theta_0$ much larger than
for the other afterglows, their linear correlation coefficient is $r (E_0, \theta_0) = 0.68 \pm
0.04$, \ie a 5\% probability of obtaining this correlation by chance.

 The $\gamma$-ray energy output $E_\gamma$ for our set of afterglows, calculated from the
$k$-corrected isotropic-equivalent ${\cal E}_\gamma$ of the GRB emission in the 20--2000 keV 
obtained by Bloom, Frail \& Sari (2001) and the jet aperture resulting from afterglow modeling,
\ie $E_\gamma = (1/2) {\cal E}_\gamma (1 - \cos \theta_0)$, are listed in Table 3. 
We find that, excluding 000418, for which the jet opening is much larger than for the other cases
analyzed here, the dynamical range of $E_\gamma$ is 30, \ie a factor 3 larger than obtained by 
Frail \etal (2001) after determining the jet aperture from the jet-break time (\eq [\ref{tjet}]).

 Table 3 also shows the resulting 20--2000 keV GRB efficiency\footnote{
  No value is given for GRB 970508, as we interpreted its afterglow emission arising from a jet 
  seen from outside its initial opening, \ie not from the same ejecta that has produced the 
  GRB emission}  
defined by $\epsilon_\gamma = E_\gamma/(E_\gamma+E_0)$. Note that, with the exception of 980519, 
$\epsilon_\gamma$ is larger than $\sim$ 50\%, most likely exceeding the ability of internal shocks 
to channel the dissipated energy into the 20--2000 keV band. 
This suggests that, during the GRB phase, jets are inhomogeneous on angular scales smaller than 
$\theta_0$, such that we are biased toward observing bursts whose outflow has a bright spot 
moving directly toward the observer or very close to the observer's line of sight toward the
GRB source (Kumar \& Piran 2000). Then the true efficiency of the GRB is much smaller and the
jet energy $E_0$ at the beginning of the afterglow phase is closer to the jet energy before
the GRB phase (Piran \etal 2001).

\subsection{External Medium}

 Our results show that models with a homogeneous medium can accommodate the broadband emission of
all EIGHT afterglows, while wind-like medium is consistent with the observations in at most four
cases. If our assumption regarding the jet uniformity is correct, then a GRB model involving a 
massive star is allowed in the remaining four cases only if there is a mechanism for homogenizing 
the wind surrounding the star prior to its interaction with the jet. Ramirez-Ruiz \etal (2001) have 
shown that the interaction between the wind of a Wolf-Rayet star and a circumstellar medium of 
$n = 1\cm3$ leads to the formation of a quasi-uniform, hot shell of density $\sim 10^3 \cm3$, 
extending from $\simg 10^{16}$ cm up to $\sim 10^{18}$ cm. More tenuous (or colder) media could 
produce thicker and less dense shells, consistent with the range of densities found here. 
  
 The particle density given in Table 2 for homogeneous media range from values typical for the
interstellar medium (970508, 980519, 990510, 010222) to those of diffuse hydrogen clouds (991208, 
991216, 000301c, 000418, 000926). In one case (990123) we find an external density below $10^{-2} 
\cm3$, characteristic of a hot component of the interstellar medium or a galactic halo. A similar 
low density was also obtained for the afterglow 980703 (PK01). These values are 2--5 orders of 
magnitude smaller than those implied by the $N_H$ column densities inferred by Galama \& Wijers 
(2001) for 970508, 980519, 980703, 990123, and 990510 from the absorption seen in their soft 
$X$-ray spectra. Furthermore, external densities higher than inferred by us are expected if GRBs 
are related with the death of massive stars, as in the collapsar model (Woosley 1993, Paczy\'nski 
1998, MacFadyen \& Woosley 1999). Our results are compatible with the above results/expectations 
if the gas in the vicinity of the GRB was evacuated prior to the jet ejection. Recently Scalo \& 
Wheeler (2001) have pointed out that the supernovae and H II region winds occurring in a cluster 
of massive stars form ``superbubbles" within giant molecular clouds, with local densities that 
range over few orders of magnitude, possibly being as low as $10^{-3} \cm3$, depending on the 
superbubble age, ambient medium and power input from supernovae.

\subsection{Jet Lorentz Factor and Mass}
\label{Gm}

 The afterglow emission is only weakly dependent on the initial jet Lorentz factor $\Gamma_0$,
which determines the evolution of the radiative losses in the early afterglow. Thus $\Gamma_0$ 
cannot be directly constrained through afterglow modelling. However, the inferred jet parameters 
can be used to determine the jet Lorentz factor $\Gamma$ during the afterglow phase:
\beq
  \Gamma \simeq 400\; \left( \frac{E_{0,50}}{\theta_{0,-1}^2 n_0} \right)^{1/8}\; 
                   \left(\frac{t}{1+z} \right)^{-3/8} \;,
\label{Gamma}
\eeq
where the usual notation $X = 10^n X_n$ was used and $t$ is measured in seconds. Thus $\Gamma_0$ 
can be calculated if one knows when the afterglow began, \ie the jet deceleration timescale $t_0$. 
In a few bursts (Giblin \etal 1999, Tkachenko \etal 2000), soft $X$-ray emission has been observed 
from the end of the GRB phase up to $10^4$ s, indicating that the external shock had already set 
in by the end of the GRB. In other cases (Pian \etal 2001, 't Zand \etal 2001), no $X$-ray emission 
has been detected after the GRB, suggesting that $t_0$ is larger then the burst duration. 
To constrain $\Gamma_0$, we assume\footnote{
 This assumption is also used for GRB 970508 although, in our interpretation of the brightening
 of its afterglow, the burst and afterglow emission arise from different ejecta. The inclusion
 of 970508 does not change the following conclusions regarding the jet initial Lorentz factor
 and mass}
that the observed GRB duration is a good measure of $t_0$.  Equation (\ref{Gamma}) shows that 
$\Gamma$ has a moderate dependence on $t$, thus the error due to this assumption is, likely, not 
too large. 

 Table 3 lists the values of $\Gamma_0$ obtained for the best fit parameters given Table 2. 
As shown in Figure 7, $\Gamma_0$ varies between 70 and 300 and is anticorrelated with the jet 
initial opening angle. Their linear correlation coefficient $r (\Gamma_0,\theta_0) = - 0.47 
\pm 0.11$ corresponds to a 20\% chance of obtaining by chance this correlation in the null 
hypothesis. The anticorrelation of $\Gamma_0$ with $\theta_0$ has been recently suggested
by Salmonson \& Galama (2002) based on the positive correlation they observed in several cases
between the GRB pulse lag-time and the afterglow jet-break time. We note, however, that the
dependence they infer between $\Gamma_0$ and $\theta_0$ ($\Gamma_0 \propto \theta_0^{-8/3}$)
is much stronger than that found by us from afterglow modeling ($\Gamma_0 \propto \theta_0^{-0.3}$,
see Figure 7).

 That wider jets have lower bulk Lorentz factors may also be the origin of the GRB pulse 
lag-time anticorrelation with the burst peak luminosity found by Norris, Marani \& Bonnell 
(2000) and Salmonson (2000). For the same kinetic energy, narrower jets have a larger energy 
per solid angle, which could lead to a higher GRB peak luminosity. If, in the comoving frame 
of the burst, the pulse duration and/or peak time dependence on photon energy are set by a 
process (\eg electron cooling) whose timescale is the same in all bursts then, due to the 
relativistic contraction of time, GRBs from faster jets would have smaller pulse lag-times. 
Then the $\Gamma_0-\theta_0$ anticorrelation implies that (narrower) jets with higher peak 
luminosities (are faster and) yield shorter $\gamma$-ray pulse lags.

 As indicated in Table 3, the product $\Gamma_0 \theta_0$ ranges from $\siml 10$ to 80, implying 
that, during the GRB phase, due to relativistic beaming, the observer receives emission from 
only a small fraction, less than 2\%, of the jet surface. Thus calculations of the jet $\gamma$-ray 
output $E_\gamma$ obtained by equating the energy per solid angle in the $\Gamma_0^{-1}$ region 
visible during GRB phase to the energy per solid angle within the much wider region seen during 
the afterglow phase could lead to rather unreliable results.

 From the jet energy $E_0$ at the beginning of the afterglow phase, one can also calculate the
initial et mass: $M_0 = c^{-2} E_0/\Gamma_0$. The results are given in Table 3 and Figure 7.
The jet mass is correlated with the jet opening, the linear correlation coefficient being
$r(\theta_0,M_0) = 0.96 \pm 0.03$. We note that $M_{jet}$ increases slower than $\theta_0^2$, 
thus the ratio between the jet mass and the stellar mass within the jet opening decreases with 
increasing $\theta_0$. For a $10\,M_\odot$ GRB progenitor and the jet masses given in Table 3,
this ratio is between $10^{-5}$ and $10^{-4}$, indicating that prior to the jet release the 
stellar material along the jet direction is strongly evacuated.

\subsection{Microphysical Parameters}

 The results of Table 2 show that the fractional energy in the magnetic field spans three orders
of magnitude and that the index $p$ of the power-law distribution of shock-accelerated electrons 
is not universal. In four of the afterglows analyzed here, the shallow fall-off of either the 
radio or the optical light-curve after the jet break requires $p \sim 1.5$. M\'esz\'aros, Rees 
\& Wijers (1998) have shown that, for a fixed $p$, variations in the jet energy per solid angle 
could lead to range of light-curve decay. Because the observer receives radiation from the entire 
jet surface after the jet-break time $t_j$, the internal structure of the jet has little effect 
on the light-curve decay index after $t_j$, thus we believe that the values of $p$ determined by 
modelling the post jet-break afterglow decay are not sensitive to the angular structure of the
outflow.

 We note that, for a fractional energy electron $\epsilon$ close to equipartition, the hard 
electron distributions ($p < 2$) identified in the 991208, 991216, 000301c, and 010222 afterglows, 
lead to a $\nu_*$-break passing through the optical band at/after few days, yielding the 
steepening seen in the optical emission of these afterglows.

\section{Conclusions}

 Our modelling of the broadband emission of ten afterglows reveals several properties of GRB jets, 
which represent constraints on the models for GRB progenitors (Woosley 1993, Paczy\'nski 1998, Vietri 
\& Stella 1998, MacFadyen \& Woosley 1999, M\'esz\'aros, Rees \& Wijers 1999, MacFadyen, Woosley \& 
Heger 2001): \\
 $1)$ the jet energy has a relatively narrow distribution, the values determined here being within 
      a factor of 5, around $\sim 5 \times 10^{50}$ erg, \\
 $2)$ the jet initial Lorentz factor is between $\sim 100$ and 300,\\
 $3)$ narrower jets are less massive and more relativistic than wider jets \\
 $4)$ the baryonic mass encountered by the jet (as it breaks out) is less than $10^{-4}$ of 
      the material that the GRB progenitor had initially within the jet aperture, \\  
 $5)$ the surrounding medium does not have, in general, the $r^{-2}$ profile expected for the 
    unperturbed wind of a massive GRB progenitor. In most cases we find that the density of 
    the external medium is between $0.1\cm3$ and $100\cm3$.

 The conclusions and the jet parameters presented here were obtained by modelling the afterglow 
data within a specific framework and under certain assumptions, the most notable being the 
uniformity of the jet and the constancy of the energy release parameters ($\epsel$, $\epsmag$).
For simplicity, the observer was located on the jet symmetry axis. Until the time when first
observations are done (few hours to 1 day), the narrow jets considered here undergo significant
lateral spreading, so that the afterglow light-curves seen by an observer located off the jet 
axis (but still within the initial jet opening, to allow the GRB to be detected and localized) 
differ little from those seen by an on-axis observer.  

 More complex jet models for GRB afterglows, such as that of a structured jet proposed by Rossi 
\etal (2001), or a hydrodynamical treatment of the jet lateral spreading (Granot \etal 2001), 
may yield different jet parameters and constraints on GRB progenitors than presented here. 
We note that the existence of a quasi-universal jet energy has also been established in a 
less model-dependent way by Piran \etal (2001), based on the narrow width of the afterglow 
$X$-ray luminosity at 1/2 day, when a good fraction of the entire jet is visible to the 
observer, thus this property should also be present in more sophisticated jet models.

\acknowledgments{AP acknowledges the supported received from Princeton University through
                 the Lyman Spitzer, Jr. fellowship.} 



\begin{table}[h]
\begin{center}
  {\bf TABLE 1.} 
  Evolution of spectral parameters and indices of power-law light-curves at $\nu > \nu_i$. \\ [4ex]
\begin{tabular}{cccc|cc|c|ccc}
 \hline \hline
 \rule[-2.5mm]{0mm}{8mm}
  $n \propto$ & $t/t_j$ & $\nu_a\propto$ & $\nu_i\propto$ & \multicolumn{2}{c|}{$\nu_c\propto$} &
           $F_p \propto$ & \multicolumn{3}{c}{$-{\rm d}\ln F_\nu/{\rm d}\ln t$}  \\
 \rule[-2mm]{0mm}{4mm}
    &  &  &  &  &  &  & $\nu_i < \nu < \nu_c$  &  \multicolumn{2}{c}{$\nu_c < \nu$}  \\
 \rule[-2mm]{0mm}{6mm}
    &  &  &  &  $Y < 1$  & $Y > 1$  &   &   &  $Y < 1$  &  $Y > 1$   \\
 \hline
 \rule[-2.5mm]{0mm}{8mm}
 $r^0$  & $ < 1$  &   $t^0$    & $t^{-3/2}$ & $t^{-1/2}$ &  $t^{(3p-8)/(8-2p)}$ &   $t^0$    &
         $\frac{3}{4}p-\frac{3}{4}$ &  $\frac{3}{4}p-\frac{1}{2}$  & $\frac{3}{4}p - \frac{1}{4-p}$  \\
\rule[-2.5mm]{0mm}{8mm}
 $r^{-2}$  & $ < 1$  & $t^{-3/5}$ & $t^{-3/2}$ & $t^{1/2}$  &  $t^{(3p-4)/(8-2p)}$ & $t^{-1/2}$ &
           $\frac{3}{4}p-\frac{1}{4}$ &  $\frac{3}{4}p-\frac{1}{2}$  & $\frac{3}{4}p - \frac{p}{8-2p}$ \\
\rule[-2.5mm]{0mm}{8mm}
 $r^0,r^{-2}$ & $>1$ & $t^{-1/5}$ & $t^{-2}$ & $t^0$ & $t^{(2p-4)/(4-p)}$ & $t^{-p}$ & 
                              $p$ & $p$ & $p-\frac{p-2}{4-p}$\\
 \hline \hline
\end{tabular}
\end{center}
\end{table}

\begin{table}
\begin{center}
 {\bf TABLE 2.} 
   Best fit parameters for a homogeneous medium and 90\% confidence levels for ten GRB afterglows. \\ [4ex]
\begin{tabular}{ccccccccc}
 \hline \hline
\rule[-2.5mm]{0mm}{8mm}
  GRB  & $E_0$ & $\theta_0$ & $n$ & $\epsel$ & $\epsmag$ & $p$ & $\chi^2$/dof & $n \propto r^{-2}$ ?  \\
         & ($10^{50}$ erg) &   (deg)    & ($\cm3$) &  ($10^{-2}$)   &          &         &       &   \\
 \hline
\rule[-2.5mm]{0mm}{8mm}
  970508 & 20 & 18.3 & 0.75 & 11 & $4.5\times 10^{-2}$ & 2.18 & 780/279 & yes \\
\rule[-2.5mm]{0mm}{8mm}
  980519$^a$ & $4.1^{+4.8}_{-1.4}$ & $2.3^{+0.2}_{-0.2}$ & $0.14^{+0.32}_{-0.03}$ & 
       $11^{+4}_{-3}$ & $(3.5^{+32}_{-2.3})\times 10^{-5}$ & $2.78^{+0.07}_{-0.04}$ & 53/46 & no \\
\rule[-2.5mm]{0mm}{8mm}
  990123 & $1.5^{+3.3}_{-0.4}$ & $2.1^{+0.1}_{-0.9}$ & $(1.9^{+0.5}_{-1.5}) \, 10^{-3}$ &
       $13^{+1}_{-4}$ & $(7.4^{+23}_{-5.9})\times 10^{-4}$ & $2.28^{+0.05}_{-0.03}$ & 55/56 & no \\
\rule[-2.5mm]{0mm}{8mm}
  990510 & $1.4^{+4.9}_{-0.5}$ & $3.1^{+0.1}_{-0.5}$ & $0.29^{+0.11}_{-0.15}$ & 
     $2.5^{+3.1}_{-0.6}$ & $(5.2^{+42}_{-4.7})\times 10^{-3}$ & $1.83^{+0.18}_{-0.01}$ & 36/69 & no \\
\rule[-2.5mm]{0mm}{8mm}
  991208 & $2.4^{+2.8}_{-0.8}$ & $12.8^{+1.5}_{-2.2}$ & $18^{+22}_{-6}$ & $5.6^{+2.1}_{-0.9}$ & 
           $(3.5^{+6.0}_{-2.1})\times 10^{-2}$ & $1.53^{+0.03}_{-0.03}$ & 112/97 & yes \\
\rule[-2.5mm]{0mm}{8mm}
  991216 & $1.1^{+1.0}_{-0.4}$ & $2.7^{+0.4}_{-1.0}$ & $4.7^{+6.8}_{-1.8}$ & $1.4^{+0.4}_{-0.3}$ & 
           $(1.8^{+3.4}_{-0.7})\times 10^{-2}$ & $1.36^{+0.03}_{-0.03}$ & 42/41 & yes \\
\rule[-2.5mm]{0mm}{8mm}
  000301c & $3.3^{+0.3}_{-0.5}$ & $13.7^{+0.6}_{-0.5}$ & $27^{+5}_{-5}$ & $6.2^{+1.4}_{-1.3}$ & 
           $(7.2^{+3.1}_{-1.5})\times 10^{-2}$ & $1.43^{+0.05}_{-0.07}$ & 119/96 & maybe \\
\rule[-2.5mm]{0mm}{8mm}
  000418 & $32^{+120}_{-14}$ & $50^{+30}_{-12}$ & $27^{+250}_{-14}$ & $7.6^{+1.2}_{-3.2}$ &
           $(6.6^{+48}_{-5.7})\times 10^{-3}$ & $2.04^{+0.07}_{-0.18}$ & 55/61 & yes \\
\rule[-2.5mm]{0mm}{8mm}
  000926 & $3.2^{+0.3}_{-0.3}$ & $8.1^{+0.5}_{-0.6}$ & $22^{+5}_{-5}$ & $10^{+2}_{-2}$ & 
           $(6.5^{+1.5}_{-1.1})\times 10^{-2}$ & $2.40^{+0.01}_{-0.02}$ & 159/102 & no  \\
\rule[-2.5mm]{0mm}{8mm}
  010222 & 5.1 & 4.6 & 1.7 & 0.43 & $6.7\times 10^{-5}$ & 1.35 & 230/87 & yes \\
 \hline \hline
\end{tabular}
\end{center}
\hspace*{1cm} $^a$ redshift unknown. $z=1$ was assumed.
\end{table}

\begin{table}
\begin{center}
 {\bf TABLE 3.} 
  Burst properties and jet characteristics inferred from the best fit parameters given in Table 2. \\ [4ex]
\begin{tabular}{cccccccc}
 \hline \hline
\rule[-2.5mm]{0mm}{8mm}
  GRB  & z &  $E_\gamma^{(b)}$ & $\epsilon_\gamma^{(c)}$ & $t_\gamma^{(d)}$ & 
              $\Gamma_0^{(e)}$ & $\Gamma_0 \theta_0$ & $M_{jet}^{(f)}$      \\ 
        &  & ($10^{50}$ erg) & & (s) & & & ($10^{-6} M_\odot$)  \\ 
 \hline
\rule[-2.5mm]{0mm}{8mm}
  970508 & 0.84     &  ...   &   ...    &   35   &   150   &  47  &  8.2   \\
\rule[-2.5mm]{0mm}{8mm}
  980519 & 1$^{(a)}$&  0.52  &   0.11   &   40   &   250   &  10  &  0.90   \\
\rule[-2.5mm]{0mm}{8mm}
  990123 & 1.60     &   4.9  &   0.76   &  100   &   300   &  11  &  0.28   \\
\rule[-2.5mm]{0mm}{8mm}
  990510 & 1.62     &   1.3  &   0.48   &  100   &   140   &  7.8 &  0.53   \\
\rule[-2.5mm]{0mm}{8mm}
  991208 & 0.71     &   18   &   0.88   &   60   &    68   &  15  &  2.0    \\
\rule[-2.5mm]{0mm}{8mm}
  991216 & 1.02     &  3.0   &   0.73   &   30   &   150   &  7.0 &  0.43   \\
\rule[-2.5mm]{0mm}{8mm}
 000301c & 2.03     &  6.6   &   0.66   &   10   &   160   &  38  &  1.2    \\
\rule[-2.5mm]{0mm}{8mm}
  000418 & 1.12     &  148   &   0.82   &   30   &    90   &  78  &   20    \\
\rule[-2.5mm]{0mm}{8mm}
  000926 & 2.07     &   15   &   0.82   &   25   &   130   &  19  &  1.4    \\
\rule[-2.5mm]{0mm}{8mm}
  010222 & 1.48     &   11   &   0.67   &  120   &   110   &  9.2 &  2.5    \\
 \hline \hline
\end{tabular}
\end{center}
\hspace*{4cm} \parbox{13cm}{ \small
 $^a$ assumed \\
 $^b$ $k$-corrected jet energy output in the 20--2000 keV band (Bloom \etal 2001)\\
 $^c$ efficiency of $\gamma$-ray emission \\
 $^d$ observed GRB duration  \\
 $^e$ jet Lorentz factor at end of GRB \\
 $^f$ jet mass \\
}
\end{table}

\clearpage

\begin{figure*}
\centerline{\psfig{figure=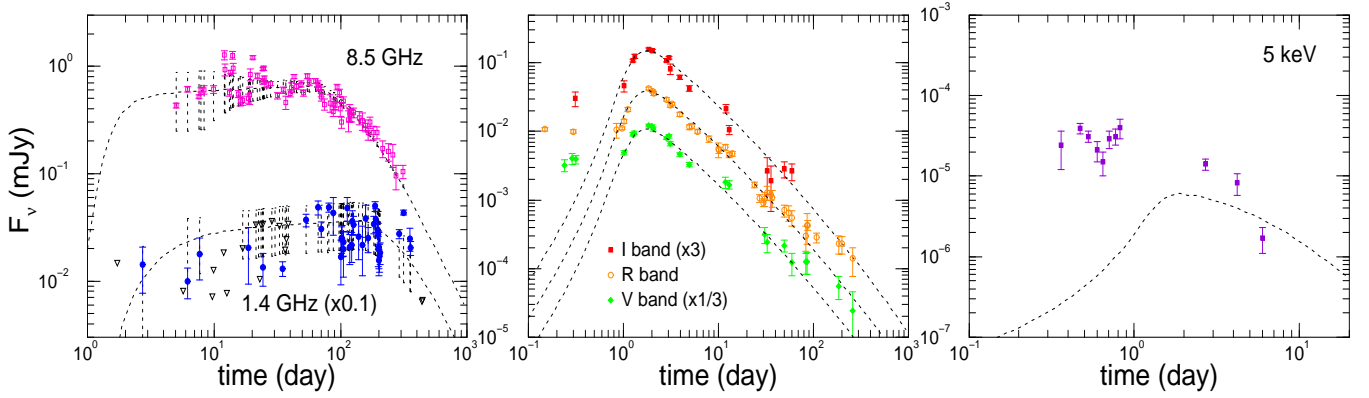,width=18cm,height=5.5cm}}   
\figcaption{
 Best fit for the afterglow of GRB 970508, obtained with a wind-like medium. Parameters are
 given in text. A jet seen by an observer located outside the jet initial opening was assumed 
 in order to obtain the brightening observed after $0.65$ days. The afterglow emission seen
 before this time was not included in the fit and could arise from some ejecta located
 outside the "central", more energetic jet. 
 The cooling frequency $\nu_c$ is below the optical range. The injection frequency $\nu_i$ 
 passes through the optical domain slightly before 1 day and through the radio domain at 
 $\sim 100$ days. The jet Lorentz factor $\Gamma$ falls below 2 at $\sim 80$ days.
 Dotted vertical lines indicate the amplitude of the interstellar scintillation, triangles
 showing 2$\sigma$ upper limits.
 The $I$ and $V$ band fluxes have been multiplied, for clarity, by the factors indicated. 
 The host galaxy contribution inferred by Zharikov \etal (1998) ($I=24.13 \pm 0.28$, 
 $R = 25.55 \pm 0.19$, $V = 25.80 \pm 0.14$) has been subtracted. 
}
\end{figure*}

\begin{figure*}
\centerline{\psfig{figure=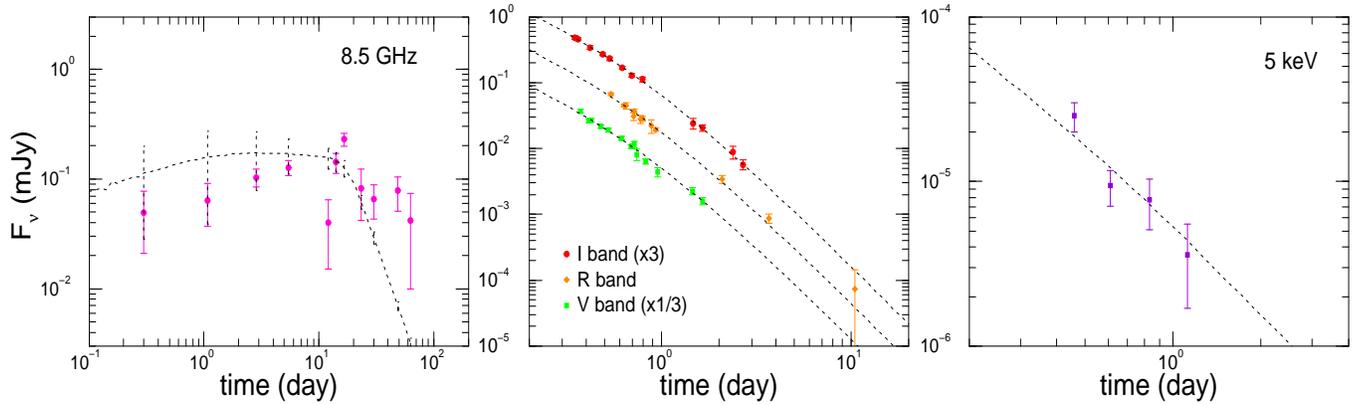,width=18cm,height=5.5cm}}
\figcaption{
 Best fit for the afterglow of GRB 980519, obtained with a homogeneous medium. 
 The model, whose parameters are given in Table 2, has $\nu_c$ between the optical and 
  $X$-ray domains. The electron cooling is due mostly to inverse Compton scatterings. 
 Optical data has been corrected for Galactic extinction of $E(B-V)=0.267$ (Jaunsen \etal 2001). 
 For this afterglow a redshift has not been measured. We have assumed $z=1$, a value 
  typical for other GRBs. 
}
\end{figure*}

\begin{figure*}
\centerline{\psfig{figure=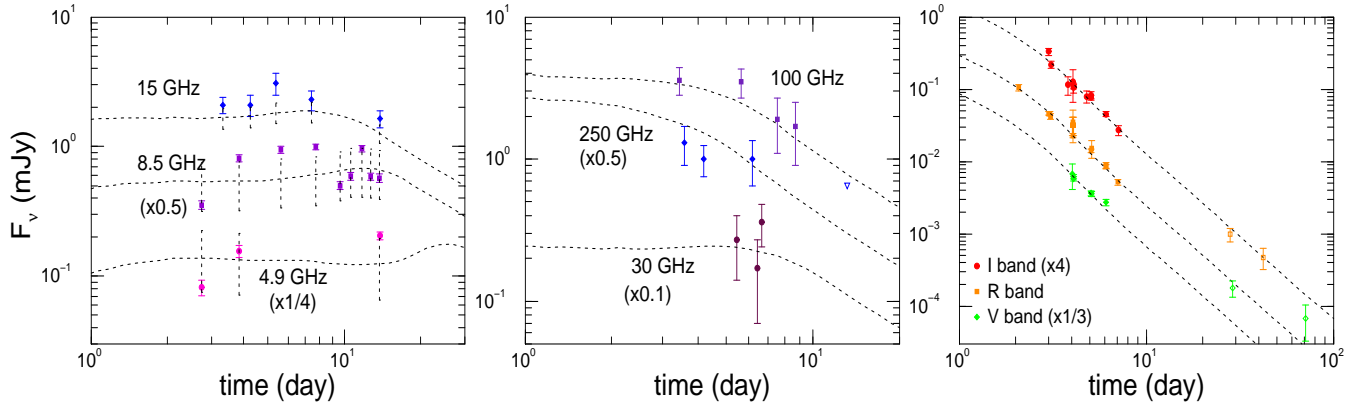,width=18cm,height=5.5cm}}
\figcaption{
 Best fit for the afterglow of GRB 991208, obtained with a homogeneous medium. 
 The jet has the parameters given in Table 2 and $t_j \siml 2$ days, thus it undergoes 
  significant lateral spreading and its edge is visible at the time of the first observations. 
  $\Gamma$ is below 4 at the time of observations. The $\nu_c$ and the cut-off frequency
  $\nu_*$ are below the optical domain. 
 The optical emission after 10 days (shown with open symbols) exceeds the model expectations, 
  suggesting the existence of a supernova contribution (Castro-Tirado \etal 2001). 
 The host galaxy contribution ($I=23.46 \pm 0.49$, $R=24.27 \pm 0.15$, $V=24.55 \pm 0.16$) 
  was subtracted. The triangle is a 2$\sigma$ upper limit on the 250 GHz flux.
}
\end{figure*}

\begin{figure*}
\centerline{\psfig{figure=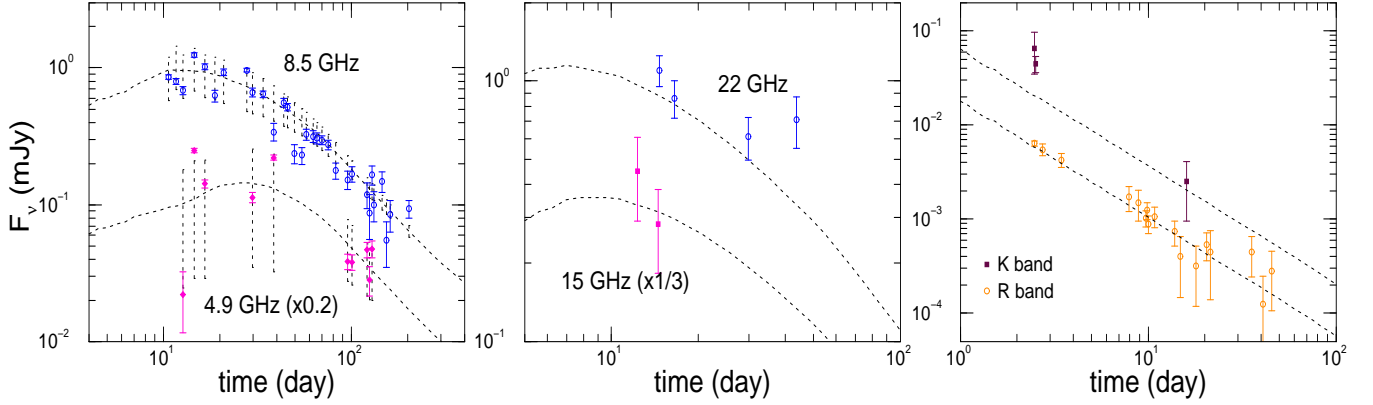,width=18cm,height=5.5cm}}
\figcaption{
 Best fit for the afterglow of GRB 000418, obtained with a homogeneous medium, and
  parameters given in Table 2. 
 The $\nu_c$ is below the optical range, the Compton parameter is above unity at all times, 
  and $\Gamma < 2$ after $\sim 10$ days. The $\nu_i$ falls below radio frequencies at $\sim 20$ 
  days, yielding the observed fall-offs at 5 and 8 GHz. 
 The host contamination of $R = 23.9 \pm 0.2$ (Metzger \etal 2000) has been subtracted
  from the reported $R$-band magnitudes.
}
\end{figure*}

\begin{figure*}
\centerline{\psfig{figure=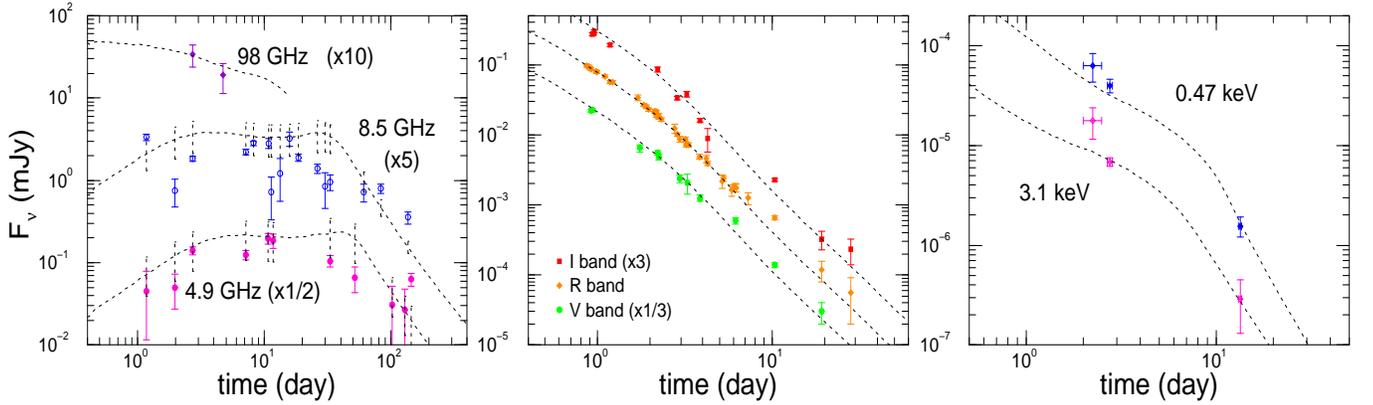,width=18cm,height=5.5cm}}
\figcaption{
 Best fit for the afterglow of GRB 000926, obtained with a homogeneous medium. 
 The parameters of the jet are listed in Table 2. 
 The $\nu_c$ lies below the optical domain and the Compton parameter is slightly below
  unity at times when the optical observations were made. 
 The model $X$-ray emission is due to inverse Compton scatterings, with a significant 
  contribution from synchrotron.
 Optical measurements have been dereddened for host extinction with an SMC-like reddening 
  curve and $A_V = 0.18$ (Fynbo \etal 2001). The contribution of a nearby galaxy, corresponding 
  to $I = 24.50 \pm 0.11$, $R = 25.19 \pm 0.17$, and $V = 26.09 \pm 0.16$ (Price \etal 2001) 
  has been subtracted from ground-based measurements. 
}
\end{figure*}

\begin{figure*}
\centerline{\psfig{figure=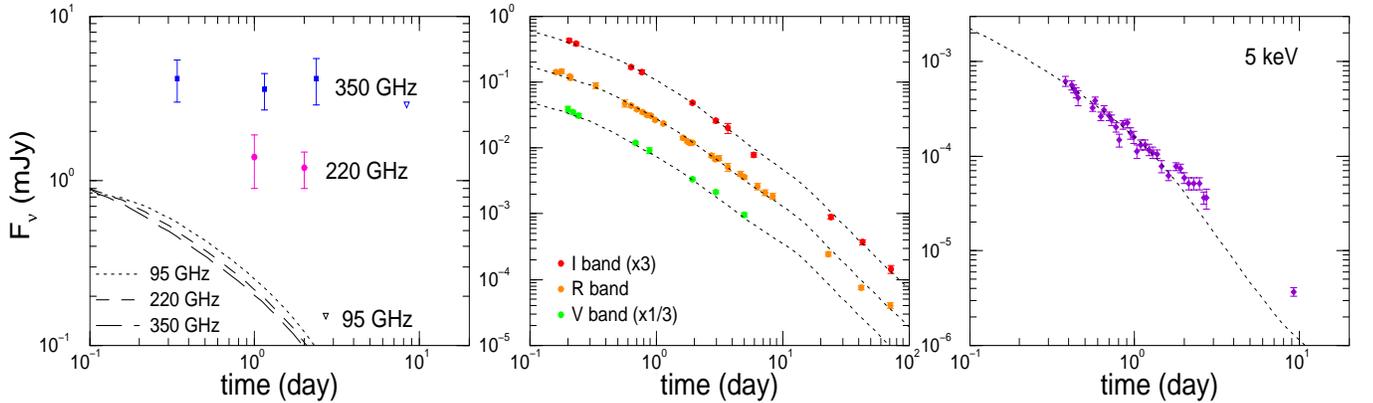,width=18cm,height=5.5cm}}
\figcaption{
 Best fit for the afterglow of GRB 010222, obtained with a homogeneous medium and parameters 
  given in Table 2, and host extinction of $A_V = 0.21$, for an assumed SMC-like reddening curve.
 The steepening seen in the optical emission at about 0.5 days is the jet-break, while that 
  at $\sim 10$ days is due to the passage of the $\nu_*$ spectral break. The electron cooling 
  is due mostly to up-scatterings. the $\nu_c$ is slightly below or within the optical domain. 
 Note that the model millimeter emission is below the constant fluxes (attributed to the host 
  galaxy) and upper limits reported by Bremer \etal (2001), Fich \etal (2001), and Kulkarni 
  \etal (2001). 
}
\end{figure*}

\begin{figure*}
\centerline{\psfig{figure=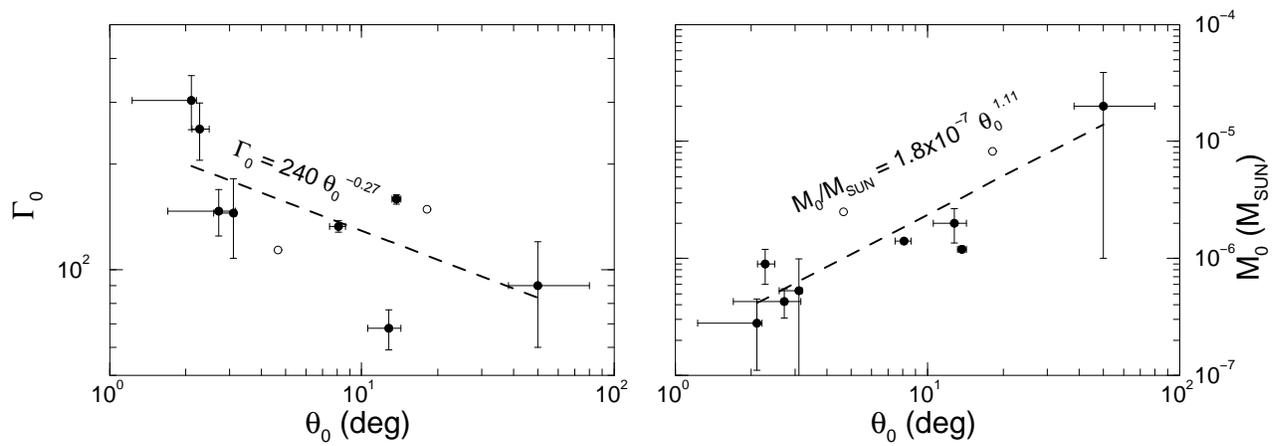}}
\figcaption{Jet Lorentz factor at the end of the GRB (see \S\ref{Gm}) and initial mass versus 
    the jet initial opening. Open symbols are for the afterglows 970508 and 010222, for which
    the best fits obtained are not satisfactory and parameter uncertainties were not determined. 
    The power-law fits to all the afterglow parameters (including 970508 and 010222) illustrate 
    that wider jets are less relativistic and more massive.}
\end{figure*}

\end{document}